\documentclass[published]{JHEP3} 

\JHEP{00(2010)000}

\usepackage{epsfig,multicol,bbm,cite}

\newcommand\fverb{\setbox\pippobox=\hbox\bgroup\verb}
\newcommand\fverbdo{\egroup\medskip\noindent%
			\fbox{\unhbox\pippobox}\ }
\newcommand\fverbit{\egroup\item[\fbox{\unhbox\pippobox}]}
\newcommand{\ud}{\mathrm{d}}
\newbox\pippobox

\def\nn{ \nonumber \\ }

\title{Hilbert Series for Flavor Invariants of the Standard Model \footnote{Imperial/TP/10/AH/06}}

\author{Amihay Hanany$^a$, Elizabeth E.~Jenkins$^b$, Aneesh V. Manohar$^b$ and Giuseppe Torri$^a$\\
$^a$Theoretical Physics Group, The Blackett Laboratory,
Imperial College London, Prince Consort Road,
London,  SW7 2AZ,  UK\\
$^b$Department of Physics, University of California at San Diego,
  La Jolla, CA 92093\\

\email{a.hanany@imperial.ac.uk,  ejenkins@ucsd.edu,}\\
\email{amanohar@ucsd.edu, giuseppe.torri08@imperial.ac.uk}

}

\received{} 		
\accepted{}		

\preprint{Imperial/TP/10/AH/06}

\abstract{The Hilbert series is computed for the lepton flavor invariants of the Standard Model with three generations including the right-handed neutrino sector needed to generate light neutrino masses via the see-saw mechanism. We also compute the Hilbert series of the quark flavor invariants for the case of four generations.}


\begin{document}

\section{Introduction}

A recent study of the lepton and quark flavor invariants was performed in the Standard Model effective theory
using methods from Invariant Theory~\cite{jm1,jm2}.  All flavor invariants and the structure of their non-trivial relations (syzygies) were found in the case of physical interest, namely the Standard Model effective theory with three generations of fermions, including a dimension-five Majorana neutrino mass operator for the light neutrinos.

The number of invariants of each degree in the fermions mass matrices is given compactly in terms of the Hilbert series for the invariants, which provides the necessary information to determine the number of independent invariants of
each degree in the basic mass matrices.
The Hilbert series for the three-generation Standard Model effective theory with a dimension-five Majorana mass term was determined in Ref.~\cite{jm2}.   The flavor invariants in the low-energy Standard Model involve the quark mass matrices $m_U$ and $m_D$, the charged lepton mass matrix $m_E$, and the light neutrino Majorana mass matrix $m_5$. The see-saw mechanism with singlet neutrinos provides an elegant way to generate the low-energy Majorana mass term using a renormalizable gauge theory. In the see-saw theory, the Majorana mass term $m_5$ is instead replaced by two other matrices, the Majorana mass matrix $M$ of the heavy singlet neutrinos, and the Dirac mass matrix coupling the singlet and doublet neutrinos, $m_\nu$.  The matrix
\begin{eqnarray}
m_5 &=& m_\nu^T M^{-1} m_\nu
\end{eqnarray}
is the Majorana mass matrix for the light neutrinos generated by integrating out the heavy singlet neutrinos. The flavor invariants in the see-saw model are considerably more complicated than in the Standard Model effective theory, and in Ref.~\cite{jm2}, the Hilbert series in the see-saw theory was obtained for two generations only. The classification of the lepton invariants for
the see-saw theory with three generations proved to be too difficult, and only a partial classification was given.
This three-generation see-saw theory is of great physical interest since it is the leading candidate theory for a renormalizable
Standard Model extension that includes neutrino masses.  

The Hilbert series for the three-generation see-saw theory can be computed using new methods developed recently in Refs.~\cite{h1,h2,h3,h4,h5} and references therein, especially the \textbf{Molien-Weyl formula}, which reduces the problem of computing the Hilbert series to that of evaluating a small number of complex integrals. This formula is a standard tool in Invariant Theory, and, given the action of a symmetry group on some mathematical objects, it allows one to compute quantities which are invariant with respect to the given symmetry.

The Molien-Weyl formula consists of two pieces: the measure of integration, which is the Haar measure of the symmetry group, and an integrand given by the so called \textbf{plethystic exponential}, which, for a function $f(t)$ vanishing at the origin, is defined as:
\begin{eqnarray}
PE[f(t)] = \exp\left( \sum^{\infty}_{r=1} \frac{f(t^r)}{r}\right).
\end{eqnarray}
The flavor invariants are symmetric functions of the masses of the leptons or of the quarks, which transform under specific representations of the flavor groups. Therefore, in order to count all the possible flavor invariants of a given theory, it is necessary to generate first all the possible symmetric products of the building blocks of these theories, \textit{i.e.} the masses of the particles. This is precisely what the plethystic exponential does: it acts like a generator for symmetrization.

Once the Molien-Weyl formula is constructed for a given theory, one can proceed by integrating over the Haar measure, thus obtaining the Hilbert series $H(t)$, which, for the cases we are interested in, is always guaranteed to be a rational function. The coefficients of the Taylor expansion of the Hilbert series give the number of flavor invariants of a certain degree.

Another interesting property of the theories we are examining is that the set of flavor invariants, together with the usual operations of sum and product have the structure of a ring which is finitely generated. Thus, it is an interesting problem to determine which are the generators of the ring and the relations among them. This is easily done by the \textbf{plethystic logarithm} of the Hilbert series which is defined as:
\begin{eqnarray}
PL[H(t)] = \sum^{\infty}_{r=1} \frac{\mu(r) \log H(t^r)}{r},
\end{eqnarray}
where $\mu(r)$ is the M\"obius function, which is defined as follows:
\begin{eqnarray}
\mu(r) := \left\{\begin{array}{lcl}
0 & & r \mbox{ has repeated prime factors}\\
1 & & r = 1\\
(-1)^n & & r \mbox{ is a product of $n$ distinct primes} \end{array}\right.
\end{eqnarray}

The first positive terms of the Taylor expansion of this function correspond to the generators of the ring of invariants and the first negative terms correspond to the relations among them (syzygies).

 In addition, we determine the quark invariants of the Standard Model for four generations of quarks as an added bonus using the same methodology.


\section{Lepton Invariants}

The Hilbert series of a general theory is defined to be
\begin{eqnarray}
H(t) &=& \sum_{r=0}^{\infty} c_r t^r,
\end{eqnarray} 
where $c_r$ is the number of invariants of degree $r$  and $c_0=1$.
Thus, the Hilbert series determines the number and degree of all flavor invariants. A general Hilbert series $H(t)$ is of the form $H(t)=N(t)/D(t)$, where the numerator $N(t)$ and the denominator $D(t)$ of the Hilbert series are polynomials in $t$. 
For the see-saw theory with three generations of leptons, the Molien-Weyl formula can be written as:
\begin{eqnarray}
 H(t)  &=& \frac{1}{(3!)^2}\prod^{3}_{k=1} \oint \frac{\ud z_k}{2 \pi i z_k}\prod^{3}_{l=1} \oint \frac{\ud w_l}{2 \pi i w_l} \prod_{i<k} (z_i - z_k)^2 \prod_{j<l} (w_j - w_l)^2  \times \nn
&& PE\Big[ \sum^{3}_{i,j} \frac{z_i}{z_j} t^2  
 + \sum^{3}_{i,j} \left(\frac{z_i}{w_j}+\frac{w_i}{z_j}\right)t + \sum^{3}_{i\leq j} \left(w_i w_j+\frac{1}{w_i w_j}\right)t\Big].
\label{e:HS33fism}
\end{eqnarray}
The numerator of the Hilbert series computed with the formula above is
\begin{eqnarray}
N(t) &=& 1+t^{4}+5 t^{6}+9 t^{8}+22 t^{10}+61 t^{12}+126 t^{14}+273
   t^{16}+552 t^{18}+1038 t^{20}\nn
   &&+1880 t^{22}+3293 t^{24}+5441 t^{26}
   +8712
   t^{28}+13417 t^{30}+19867 t^{32}+28414 t^{34}+39351 t^{36}\nn
   &&+52604
   t^{38}+68220 t^{40}+85783 t^{42}+104588 t^{44}+123852 t^{46}
   +142559
   t^{48}+159328 t^{50}\nn
   &&+173201 t^{52}+183138 t^{54}+188232 t^{56}+188232
   t^{58}+183138 t^{60}+173201 t^{62}+159328 t^{64}\nn
   &&
   +142559 t^{66}+123852
   t^{68}+104588 t^{70}+85783 t^{72}+68220 t^{74}+52604 t^{76}+39351
   t^{78}\nn
   &&+28414 t^{80}+19867 t^{82}
   +13417 t^{84}+8712 t^{86}+5441
   t^{88}+3293 t^{90}+1880 t^{92}+1038 t^{94}\nn
   &&+552 t^{96}+273 t^{98}+126
   t^{100}+61 t^{102}+22 t^{104}+9 t^{106}+5 t^{108}+t^{110}+t^{114},
   \label{num}
\end{eqnarray}
which is of degree $d_N = 114$ and palindromic, i.e. $t^{d_N} N(1/t)=N(t)$.
The denominator is
\begin{eqnarray}  
D(t) &=&   \left(1-t^2\right)^3 \left(1-t^4\right)^4 \left(1-t^6\right)^4
   \left(1-t^8\right)^2 \left(1-t^{10}\right)^2 \left(1-t^{12}\right)^3
   \left(1-t^{14}\right)^2 \left(1-t^{16}\right), \nn
   \label{den}
\end{eqnarray}
which is of degree $d_D = 162$.  
The plethystic logarithm of this Hilbert series can be written as:
\begin{eqnarray}
PL[H(t)] &=& 3 t^2 + 5 t^4 + 9 t^6 + 10 t^8 + 19 t^{10} + 40 t^{12} + 66 t^{14} + 92 t^{16} + 70 t^{18} - O(t^{20})\nn
\end{eqnarray}
There are three mass matrices in the lepton sector of the see-saw model,
$m_\nu$, $m_E$ and $M$, and their complex conjugates ${m_\nu}^\dagger$, ${m_E}^\dagger$ and $M^\dagger = M^*$.
The Dirac mass matrix of the light neutrinos $m_\nu$ and the mass matrix of the charged leptons $m_E$ are complex
$3 \times 3$ matrices, whereas the Majorana mass matrix of the heavy singlet neutrinos is a complex symmetric $3 \times 3$ matrix.  Thus, the dimension of the vector space on which the flavor symmetry transformations act is $\dim V=48$, since the complex $3 \times 3$ matrices $m_\nu$ and $m_E$ each contribute $18$ and the complex symmetric matrix $M$ contributes $12$.  Knop's theorem~\cite{knop} states that
\begin{eqnarray}
\dim V \ge d_D - d_N \ge p ,
\end{eqnarray}  
where $p$ is the number of parameters.
The number of parameters in the three mass matrices in the see-saw theory for three generations of leptons
is $p=21$, which consists of 9 masses, 6 angles and 6 phases.
Thus, Knop's theorem gives $48 \ge 48 \ge 21$.

From the point of view of Invariant Theory, computing flavor invariants in the Standard Model is equivalent to computing chiral gauge invariant operators of supersymmetric gauge theories. In particular, the Hilbert series determined from (\ref{e:HS33fism}) is the same as the one for counting gauge invariant operators in an $\mathcal{N}=1$ supersymmetric theory with gauge group $U(3) \times U(3)$, with one field transforming in the adjoint representation of the first group, corresponding to $m_E$, one bi-fundamental field and its complex conjugate, corresponding to $m_\nu$ and $m^\dagger_\nu$, and one field transforming in the rank-2 symmetric representation of the second gauge group and its complex conjugate, corresponding to $M$ and $M^\dagger$. In addition, the superpotential is set to 0.

From this perspective, the number of parameters of the see-saw theory corresponds to the dimension of the vacuum moduli space of the gauge theory, which can be elegantly determined with an argument based on the Higgs mechanism. In the Higgs mechanism, a massless vector multiplet becomes massive by `absorbing' degrees of freedom from a chiral multiplet. The gauge theory we are examining contains 9 degrees of freedom coming from the adjoint field, 18 coming from the bi-fundamentals and 12 coming from the symmetric rank-2 tensors. The vector multiplets becoming massive take away 18 degrees of freedom (9 per gauge group), leaving a total of 21, which correspond to the dimension of the moduli space of the gauge theory. Note that this counting correctly reproduces the number of parameters of the see-saw theory.

\section{Quark Invariants for Four Generations}

The invariants and relations for three generations of quarks were determined previously in Ref.~\cite{jm2}. Here we give
the Hilbert series for four generations of quarks, which is considerably more complicated than the three-generation case.
The general
counting of physical parameters of the up- and down-quark mass matrices, $m_U$ and $m_D$, respectively,
is given in Refs.~\cite{jm1,jm2}.  In the case of four quark generations, the parameters consist of 8 masses, 6 angles 
and 3 phases. 

The Hilbert series for four generations of quarks is
\begin{small}
\begin{eqnarray}
H(t) &=& \frac{1+t^{10}+3 t^{12}+2 t^{14}+4 t^{16}+4 t^{18}+2 t^{20}+3 t^{22}+3 t^{24}+2 t^{26}+4
   t^{28}+4 t^{30}+2 t^{32}+3 t^{34}+t^{36}+t^{46}}{
   \left(1-t^2\right)^2 \left(1-t^4\right)^3 \left(1-t^6\right)^4 \left(1-t^8\right)^6
   \left(1-t^{10}\right) \left(1-t^{12}\right)}.\nn
\end{eqnarray}
\end{small}
The degree of the numerator is $d_N = 46$, whereas the degree of the denominator is $d_D = 110$. 
The plethystic logarithm of this Hilbert series can be written as:
\begin{eqnarray}
PL[H(t)] &=& 2 t^2 + 3 t^4 + 4 t^6 + 6 t^8 + 2 t^{10} + 4 t^{12} + 2 t^{14} + 4 t^{16} + 4 t^{18} + t^{20} - O(t^{24}) \nn
\end{eqnarray}
There are two $4 \times 4$ matrices $m_U$ and $m_D$ and their complex conjugates, so $\dim V=64$.
From the denominator, one sees that there are $p=17$ parameters, as expected.  Knop's theorem
gives $64 \ge 64 \ge 17$.

The multi-graded Hilbert series, where $t_u$ and $t_d$ count powers of $m_U$ and $m_D$ respectively,  is found to be $H(t_u,t_d)=N(t_u,t_d)/D(t_u,t_d)$, where
\begin{eqnarray}
N(t_u,t_d) &=& \left(1+t_u^2 t_d^2\right) \left(1-t_u^2 t_d^2+t_u^4 t_d^4\right) \bigl(1-t_u^{2} t_d^4 - t_u^4
   t_d^{2}+t_u^4 t_d^6+t_u^4 t_d^8+t_u^6 t_d^4+2 t_u^6 t_d^6\nn
   &&+t_u^6 t_d^8+t_u^8 t_d^4+t_u^8 t_d^6-t_u^8 t_d^{10}-t_u^{10} t_d^8+t_u^{12} t_d^{12}\bigr),\nn
D(t_u,t_d) &=&
\left(1-t_u^2\right) \left(1-t_u^4\right) \left(1-t_u^6\right) \left(1-t_u^8\right) \left(1-t_d^2\right)
   \left(1-t_d^4\right) \left(1-t_d^6\right) \left(1-t_d^8\right) \left(1-t_u^2 t_d^2\right) \nn
   &&\left(1-t_u^4
   t_d^2\right)^2  \left(1-t_u^2 t_d^4\right)^2 \left(1-t_u^4 t_d^4\right)^2
\left(1-t_u^6 t_d^2\right)   \left(1-t_u^2 t_d^6\right),
\label{multi}
\end{eqnarray}
which is considerably more complicated than the Hilbert series of the three-flavor case found in Ref.~\cite{jm2}. The multi-graded version of Eqs.~(\ref{num},\ref{den}) has also been computed, but the result is too complicated to give here; the numerator has over 6000 terms.
The plethystic logarithm of this multi-graded Hilbert series can be written as:
\begin{eqnarray}
PL[H(t_u, t_d)] &=& t_d^2 + t_u^2 + t_d^4 + t_d^2 t_u^2 + t_u^4 + t_d^6 + t_d^4 t_u^2 + t_d^2 t_u^4 + t_u^6 + t_d^8 + t_d^6 t_u^2 + 2 t_d^4 t_u^4\nn
&& + t_d^2 t_u^6 + t_u^8 + t_d^6 t_u^4 + t_d^4 t_u^6 + t_d^8 t_u^4 + 2 t_d^6 t_u^6 + t_d^4 t_u^8 + t_d^8 t_u^6 + t_d^6 t_u^8 + t_d^{10} t_u^6\nn
&& + 2 t_d^8 t_u^8 + t_d^6 t_u^{10} + t_d^{12} t_u^6 + t_d^{10} t_u^8 + t_d^8 t_u^{10} + t_d^6 t_u^{12} +  t_d^{10} t_u^{10}  - O(t_u^{12})O(t_d^{12})\nn
\end{eqnarray}

\section{Conclusions}

In summary, we have used the methods of Refs.~\cite{h1,h2,h3,h4,h5} to find the Hilbert series for the lepton invariants of
the three-generation see-saw theory.  This problem had proven to be computationally too difficult previously, and so only
partial results were given in Refs.~\cite{jm1,jm2}.  The same techniques also were used in this work
to find the Hilbert series for the quark invariants of the Standard Model in the case of four generations. 

We would like to thank H.~Fritzsch, W.~Plessas and M.~Shifman for organizing the 2010 Oberw\"olz workshop on QCD and Strings, where this work was started. We would also like to thank John Davey and Ennio Salvioni for their kind help.

\end{document}